\documentclass[12pt]{article}
\usepackage{graphicx,color,epsfig,rotate}
\title{
Ground-states of the three-dimensional Falicov-Kimball model}
\author{P. Farka\v sovsk\'y, H. \v{C}en\v{c}arikov\'a and 
N. Toma\v{s}ovi\v{c}ov\'a\\
Institute  of  Experimental  Physics,  Slovak   Academy   of
Sciences\\
Watsonova 47, 043 53 Ko\v {s}ice, Slovakia}
\date{}
\begin{document}
\baselineskip=24pt
\maketitle

\begin{abstract}
The systematic study of ground-state properties of the three-dimensional
Falicov-Kimball model is performed by a well-controlled numerical method. 
The results obtained are used to categorize the ground-state configurations
according to common features for weak, intermediate and strong interactions.
It is shown that only a few configuration types form the basic structure of
the phase diagram. In particular, the largest regions of stability
correspond to phase segregated configurations, striped configurations and
configurations in which electrons are distributed in diagonal planes with
incomplete chessboard  structure. Near half-filling, mixtures of two phases
with complete and incomplete chessboard structure are determined. The
relevance of these results for a description of real material is discussed.
\end{abstract}

\newpage
\section{Introduction}
The Falicov-Kimball model~(FKM) has become, since its
introduction~\cite{Falicov} in 1969, one of the most popular examples
of a system of interacting electrons with short-range interactions.
It has been used in the literature to study a great variety of many-body
effects in rare-earth compounds, of which metal-insulator transitions, 
mixed-valence phenomena, and charge-density waves are the most common
examples~\cite{Chomski}. The model is based on the coexistence of two
different types of electronic states in a given material:
localized, highly correlated ionic-like states and extended,
uncorrelated, Bloch-like states. It is generally accepted that
the above mentioned cooperative phenomena result from a change in the 
occupation numbers of these electronic states, which remain themselves 
basically unchanged in their character. Taking into account only the
intra-atomic Coulomb interaction between the two types of states,
the Hamiltonian of the spinless FKM can be written as the sum of three 
terms:

\begin{equation}
H=\sum_{ij}t_{ij}d^+_id_j+U\sum_if^+_if_id^+_id_i+E_f\sum_if^+_if_i,
\end{equation}
where $f^+_i$, $f_i$ are the creation and annihilation
operators  for an electron in  the localized state at
lattice site $i$ with binding energy $E_f$ and $d^+_i$,
$d_i$ are the creation and annihilation operators
of the itinerant spinless electrons in the $d$-band
Wannier state at site $i$.

The first term of (1) is the kinetic energy corresponding to
quantum-mechanical hopping of the itinerant $d$ electrons
between sites $i$ and $j$. These intersite hopping
transitions are described by the matrix  elements $t_{ij}$,
which are $-t$ if $i$ and $j$ are the nearest neighbours and
zero otherwise (in the following all parameters are measured
in units of $t$). The second term represents the on-site
Coulomb interaction between the $d$-band electrons with density
$n_d=N_d/L=\frac{1}{L}\sum_id^+_id_i$ and the localized
$f$ electrons with density $n_f=N_f/L=\frac{1}{L}\sum_if^+_if_i$,
where $L$ is the number of lattice sites. The third  term stands
for the localized $f$ electrons whose sharp energy level is $E_f$.

Since in this spinless version of the FKM
without hybridization  the $f$-electron occupation
number $f^+_if_i$ of each site $i$ commutes with
the Hamiltonian (1), the $f$-electron occupation number
is a good quantum number, taking only two values: $w_i=1$
or 0, according to whether or not the site $i$ is occupied
by the localized $f$ electron.

Now the Hamiltonian (1) can be written as

\begin{equation}
H=\sum_{ij}h_{ij}d^+_id_j+E_f\sum_iw_i,
\end{equation}
where $h_{ij}(w)=t_{ij}+Uw_i\delta_{ij}$.

Thus for a given $f$-electron configuration
$w=\{w_1,w_2 \dots w_L\}$ defined on the three-di\-men\-sional
lattice with periodic boundary conditions, the Hamiltonian (2)
is the second-quantized version of the single-particle
Hamiltonian $h(w)=T+UW$, so the investigation of
the model (2) is reduced to the investigation of the
spectrum of $h$ for different configurations of $f$ electrons.

Despite its relative simplicity and an impressive research activity in the
past, the properties of this model remained unclear for a long time. The
crucial break in this direction has been done recently by exact
analytical~\cite{Gru,Free1,Free2} and numerical~\cite{Watson,Fark1} 
calculations. These calculations showed that the spinless 
FKM can describe (at least qualitatively) such
important phenomena observed experimentally in some rare-earth and
transition metal compounds like the discontinuous valence and metal insulator
transitions, phase separation, charge ordering, stripes formation, etc. In
addition, it was found~\cite{Fark3} that at non-zero temperatures the model 
is able to provide the qualitative explanation for the anomalous large values 
of the specific heat coefficient and for the extremely large changes of the
electrical conductivity found in some intermediate valence compounds (e.g.,
in $SmB_6$). These results indicate that the spinless FKM, in spite of its
simplicity, could be a convenient microscopic model for a description of
ground-state, thermodynamic and transport properties of real materials.
However, real materials are usually three dimensional while the most of
above mentioned results have been obtained for the limiting cases of $D=1,
D=2$ and $D= \infty$. Thus one can ask if these results, or at least some of
them hold also in three dimensions. This is the question that we would like
to answer in this paper. Here we focus our attention on the ground-state
properties of model. The special attention is devoted to examine the three 
dimensional analogs of phase segregation, charge
ordering, stripes formation and metal-insulator transitions
observed in $D=1$ and $D=2$. From this point of view the paper represents
the first attempt to describe systematically the ground-state properties of
the FKM in three dimensions. To attain this goal we use a well-controlled
numerical method that we have elaborated recently~\cite{Fark4}. The method 
is based on the simple modification of the exact diagonalization method 
on finite clusters and consists of following steps.
(i) Chose a trial configuration $w=\{w_1,w_2 \dots w_L\}$.
(ii) Having $w$, $U$ and $E_f$ fixed, find
all eigenvalues $\lambda_k$ of $h(w)=T+UW$. (iii) For a given
$N_f=\sum_iw_i$ determine the ground-state energy
$E(w)=\sum_{k=1}^{L-N_f}\lambda_k+E_fN_f$ of a particular
$f$-electron configuration $w$ by filling in the lowest
$N_d=L-N_f$ one-electron levels
(here we consider only the case $N_f+N_d=L$,
which is the point of the special interest for valence
and metal-insulator transitions caused by promotion of electrons
from localized $f$ orbitals $(f^n \to f^{n-1})$ to the conduction
band states).
(iv) Generate a new configuration $w'$ by moving a randomly
chosen electron to a new position which is chosen also as random.
(v) Calculate the ground-state energy $E(w')$. If $E(w')<E(w)$
the new configuration is accepted, otherwise $w'$ is rejected.
Then the steps (ii)-(v) are repeated until the convergence
(for given $U$ and $E_f$ ) is reached.
Of course, one can move instead of one electron (in step (iv))
simultaneously two or more electrons, thereby the convergence
of method is improved. Indeed, tests that we have performed for a wide range 
of the model parameters showed that the latter implementation of the method, 
in which  $1 < p < p_{max}$ electrons ($p$ should be chosen at random) are 
moved to new positions overcomes better the local minima of the ground state 
energy. In this paper we perform calculations with $p_{max}=N_f$. The main 
advantage of this implementation is that in any iteration step the system
has a chance to lower its energy (even if it is in a local minimum), 
thereby the problem of local minima is strongly reduced (in principle, the
method becomes exact if the number of iteration steps goes to infinity). On 
the other hand a disadvantage of this selection is that the method converges 
slower than for $p_{max}=2$ and $p_{max}=3$. To speed up the convergence of 
the method (for $p_{max}=N_f$) and still to hold its advantage we generate 
instead the random number $p$ (in step (iv)) the pseudo-random number $p$ 
that probability of choosing decreases (according to the power law) with 
increasing $p$. Such a modification improves considerably the convergence 
of the method. Repeating this procedure for different
values of $E_f$ and $U$ one can immediately study the dependence of the
$f$-electron occupation number $N_f=\sum_iw^{min}_i$ on
the $f$-level position $E_f$ (valence transitions) or the phase
diagram of the model in the $n_f-U$ plane.
This method was first used in our recent paper~\cite{Fark4} to study
the ground-state properties of the one and two-dimensional FKM. 
It was found that for small and intermediate clusters,
where the exact numerical solution is possible ($L\sim 30$), 
the method is able to reproduce exactly the ground states of the 
spinless FKM, even after relative small number of iterations 
(typically 10000 per site).
\section{Results and discussion}
To examine ground-state properties of the spinless FKM in three dimensions
we have performed an exhaustive numerical study of the model for weak
($U=1$), intermediate ($U=2$) and strong ($U=8$) interactions. For each
selected value of $U$ and $N_f$ ($N_f=0,1,..,L$) the ground-state
configuration $w^{min}$ is determined by the above described method 
(we remember that the total filling is fixed at 1). To reveal the 
finite-size effects numerical calculations were done on two different
clusters of $4\times 4 \times 4$ and $6\times 6\times 6$ sites. A direct
comparison of numerical results obtained on $4\times 4 \times 4$ and 
$6\times 6\times 6$ clusters showed that the ground-state configurations
fall into several different categories which stability regions are
practically independent of $L$. Let us start a discussion of our results
with a description of these configuration types for different 
values of $U$ and $N_f$ (in the remainder of the paper the values 
of $N_f$ always correspond to $6 \times 6\times 6$ cluster).

The largest number of configuration types is observed in the weak-coupling
limit. Going with $N_f$ from zero to half-filling ($N_f=L/2$) we have
observed the following configuration types for $U=1$. At low $f$-electron
concentrations the ground-states are the phase segregated configurations
($f$-electrons clump together while remaining part of lattice is free of
$f$-electrons) listed in Fig.~1a for two selected values of $N_f$. Since
the ground-states corresponding to the segregated configurations are 
metallic~\cite{Gruber} we arrive at an important conclusion, and namely, that 
the metallic domain that exists in the one and two dimensional FKM persists 
also in three dimensions. In the one dimensional case the region of stability 
of this metallic domain was restricted to low $f$-electron concentrations 
$n_f<1/4$ and small Coulomb interactions $U \leq 1$~\cite{Fark4,Gruber}. 
The numerical calculations performed in two dimensions revealed~\cite{Fark4} 
that with  increasing dimension the stability region of this metallic 
domain shifts to higher values of $U$ ($U\sim 3$). From this
point of view it is interesting to examine if this trend holds also for
three dimensions. To verify this conjecture we have determined the
ground-state configurations for increasing $U$ at low $f$-electron
concentrations on $4\times 4 \times 4$, $6\times 6 \times 6$ and 
$8\times 8 \times 8$ clusters. We have found that the metallic region in
$D=3$ extends up to $U\sim 5$, what confirms the trend conjectured from two
dimensional calculations (in addition, in accordance with
two-dimensional results we have found that the critical value of $U_c$ 
decreases with increasing $n_f$).
It should be noted that this result is crucial for
description of insulator-metal transitions in real materials (like 
rare-earth and transition metal compounds). In these materials the values 
of the interaction constant $U$ are much larger than the values of hopping 
integrals $t_{ij}$~\cite{Falicov}, and thus for the correct description
of valence and metal-insulator transitions in these compounds
one has to take the limit $U>t$ and not $U<t$. 
On the other hand it should be mentioned that in the Falicov-Kimball 
picture it is possible to get the metal-insulator transition much easier, 
for example by including spins. Indeed, numerical calculations performed 
for the spin-one-half FKM showed~\cite{Fark5} that the metallic domain is 
stable in this model for a wide range of model parameters, including 
large values of $U$ and $n_f$. Above the region of phase
segregation we have observed the region of stripes formation
($N_f=10,..,20$). In this region the $f$-electrons form the one-dimensional
charge lines (stripes) that can be  perpendicular or parallel (see Fig.~1b).
This result shows that the crucial mechanism leading to the stripes
formation in strongly correlated systems should be the competition between
the kinetic and short-range Coulomb interaction. Going with $N_f$ to higher
values of $N_f$ the stripes vanish and again appear at $N_f=26$, however in
a fully different distribution (see Fig.~2a). While at smaller values of $N_f$ 
the stripes have been distributed inhomogeneously (only over one half of 
lattice) the stripes in the region $N_f=26,..,31$ are distributed regularly. 
Above this region a new type of configurations (see Fig.~2b) starts to develop. 
We call them the diagonal charge planes with incomplete chessboard structure, 
since the $f$-electrons prefer to occupy the diagonal planes with 
slope 1 and within these planes they form the chessboard structure. 
Of course, there is a considerable freedom in categorization of ground state 
configurations according to some common features and the case of diagonal 
planes used by us is only one of possible ways.
The region of diagonal charge planes is relatively broad and
extends up to $N_f \sim 50$. Then follows the region in which the chessboard
structure starts to develop. As illustrated in Fig.~3a the $f$-electrons
begin to occupy preferably the sites of sublattice A, leaving the sublattice
B free of $f$-electrons. In addition, the configurations that can be considered
as mixtures of previous configuration types are also observed in this region
(see Fig.~3b). However, with increasing $N_f$ the configurations of
chessboard type become dominant. Analysing these configurations we have
found that the transition to the purely  chessboard configuration realizes
through several steps. The first step, the formation of the chessboard
structure has been illustrated in Fig.~3a. The second step is shown in
Fig.~4a. It is seen that the chessboard structure is fully developed in some
regions (planes) that are separated by planes with incomplete developed 
chessboard structure. Such a type of distribution is replaced for larger 
values of $N_f$ by a new type of distributions (step three), where both 
regions with complete and incomplete chessboard structure have 
the three-dimensional character (see Fig.~4b). 

The same picture we have observed also for intermediate values of
Coulomb interactions ($U=2$). The larger values of $U$ only slightly modify
the stability regions of some phases, but no new configuration types appear.
In particular, the domain of phase segregation, as well as the domain of
stripes formation are reduced while the domain of diagonal planes with
chessboard structure increases. This trend is observed also for larger
values of $U$. In the strong coupling limit ($U=8$) the phase segregated and
striped phases absent and the region of stability the diagonal planes extends
to relatively small values of $N_f \sim 20$. Below this value a homogeneous
distribution of $f$-electrons is observed. Thus we can conclude that all
fundamental results found in one and two-dimensional solutions of the FKM
(the phase segregation, the stripes formation, the phase separation, etc.)
holds also in three dimensions, thereby the FKM becomes interesting for a
description of ground-state properties (e.g., valence and metal-insulator
transitions induced by doping and pressure) of real (three dimensional) 
systems~\cite{Wachter}. The work in this direction is currently in progress.

In summary, the ground-state properties of the three-dimensional FKM were 
examined by a well-controlled numerical method. The results obtained were 
used to categorize the ground-state configurations according to common 
features for weak ($U=1$), intermediate ($U=2$) and strong interactions
($U=8$). It was shown that only a few configuration types form the basic 
structure of the phase diagram in the $n_f-U$ plane. In particular, the 
largest regions of stability correspond to phase segregated configurations, 
striped configurations and configurations in which electrons are distributed 
in diagonal planes with incomplete chessboard structure. Near half-filling, 
mixtures of two phases with complete and incomplete chessboard structure 
were determined.

\vspace{0.5cm}
This work was supported by the Slovak Grant Agency VEGA
under grant No. 2/4060/04 and the Science and Technology Assistance
Agency under Grant APVT-20-021602. Numerical results were obtained using
computational resources of the Computing Centre of the Slovak 
Academy of Sciences.


\newpage
\begin{figure}[hb]
\hspace{1cm}
\includegraphics[angle=0,width=14cm,scale=1]{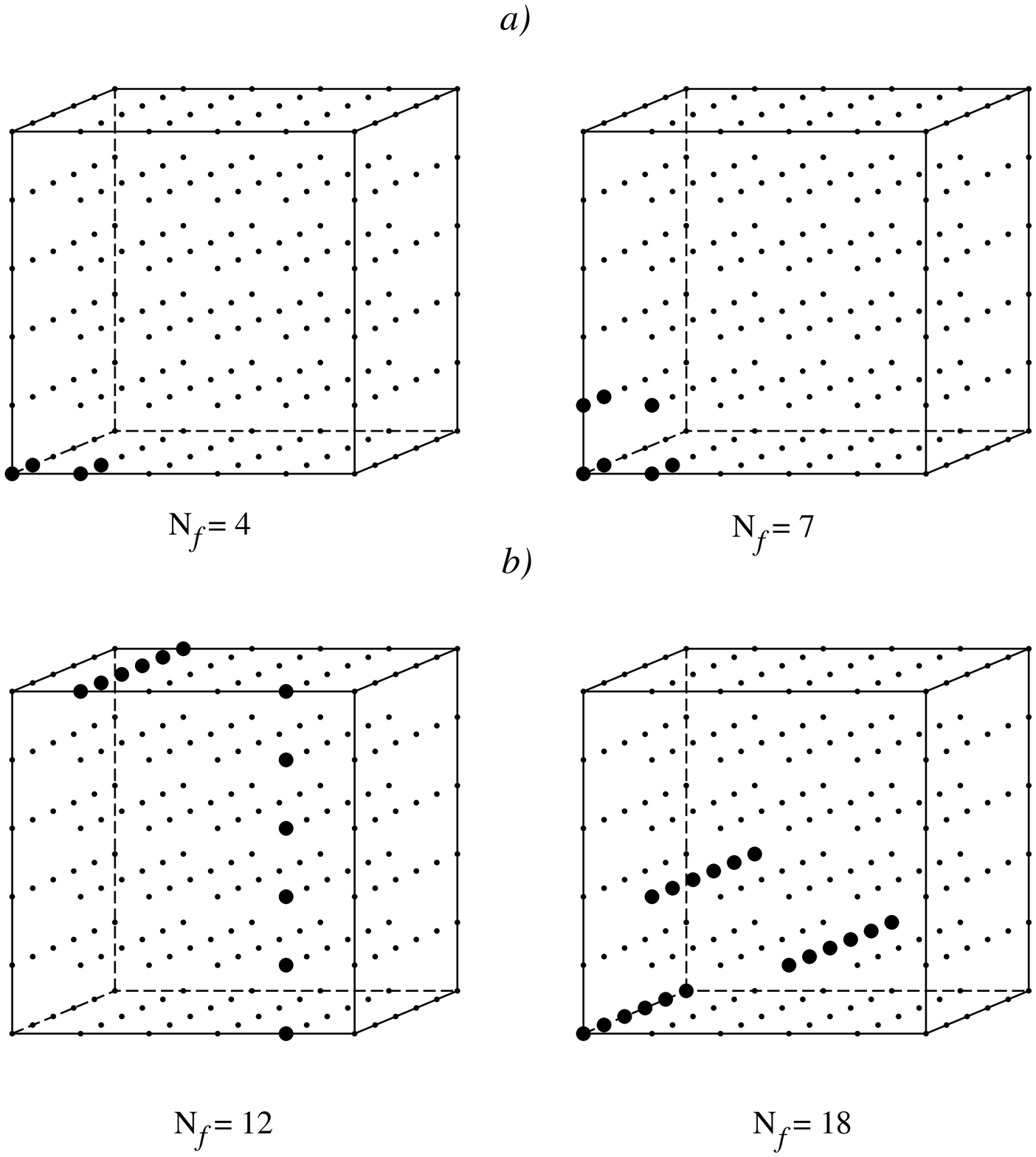}
\caption{ 
Typical examples of phase segregated (a) and striped (b) configurations 
obtained for $U=1$ and $L=6\times 6\times 6$. Large dots: occupied sites; 
small dots: vacant sites.
}
\label{fig1}
\end{figure}

\newpage
\begin{figure}[hb]
\hspace{1cm}
\includegraphics[angle=0,width=14cm,scale=1]{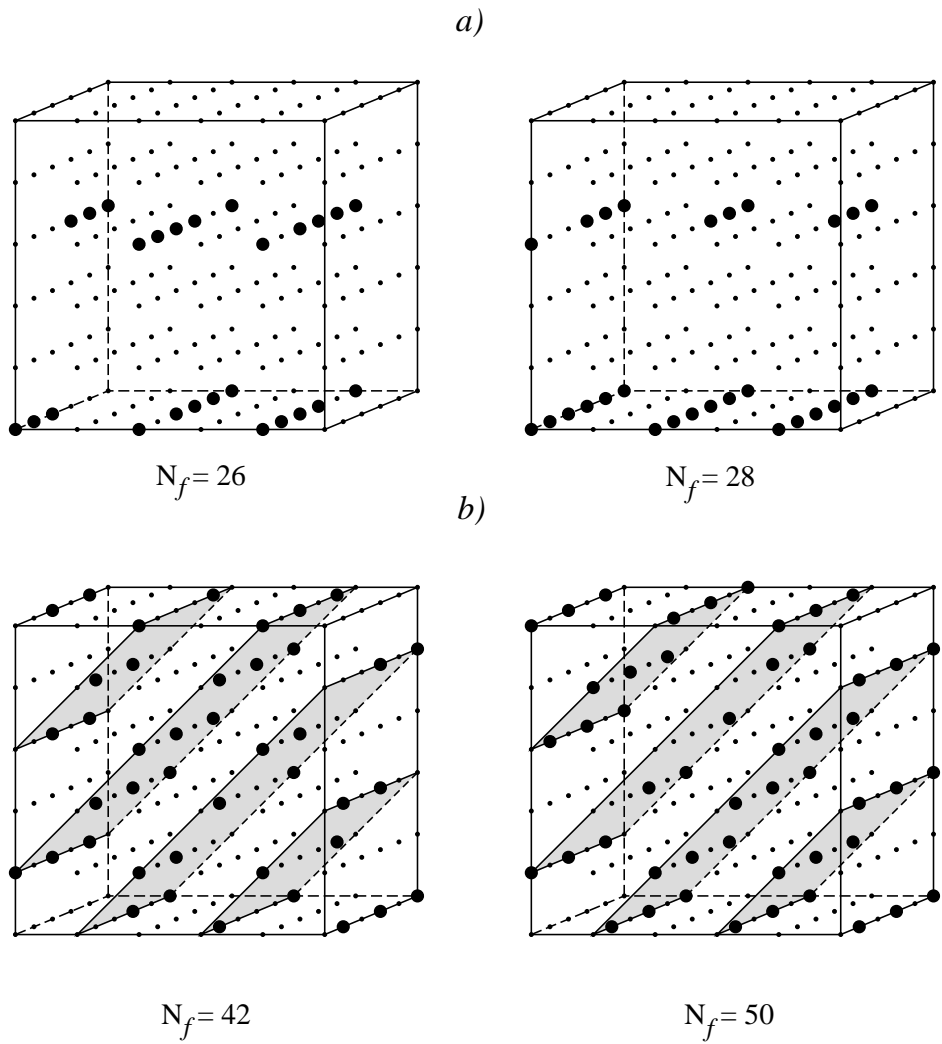}
\caption{ 
Typical examples of striped configurations with regular distribution 
(a) and diagonal charge planes with an incomplete chessboard structure
(b)  obtained for $U=1$ and $L=6\times 6\times 6$.
}
\label{fig2}
\end{figure}

\newpage
\begin{figure}[hb]
\hspace{1cm}
\includegraphics[angle=0,width=14cm,scale=1]{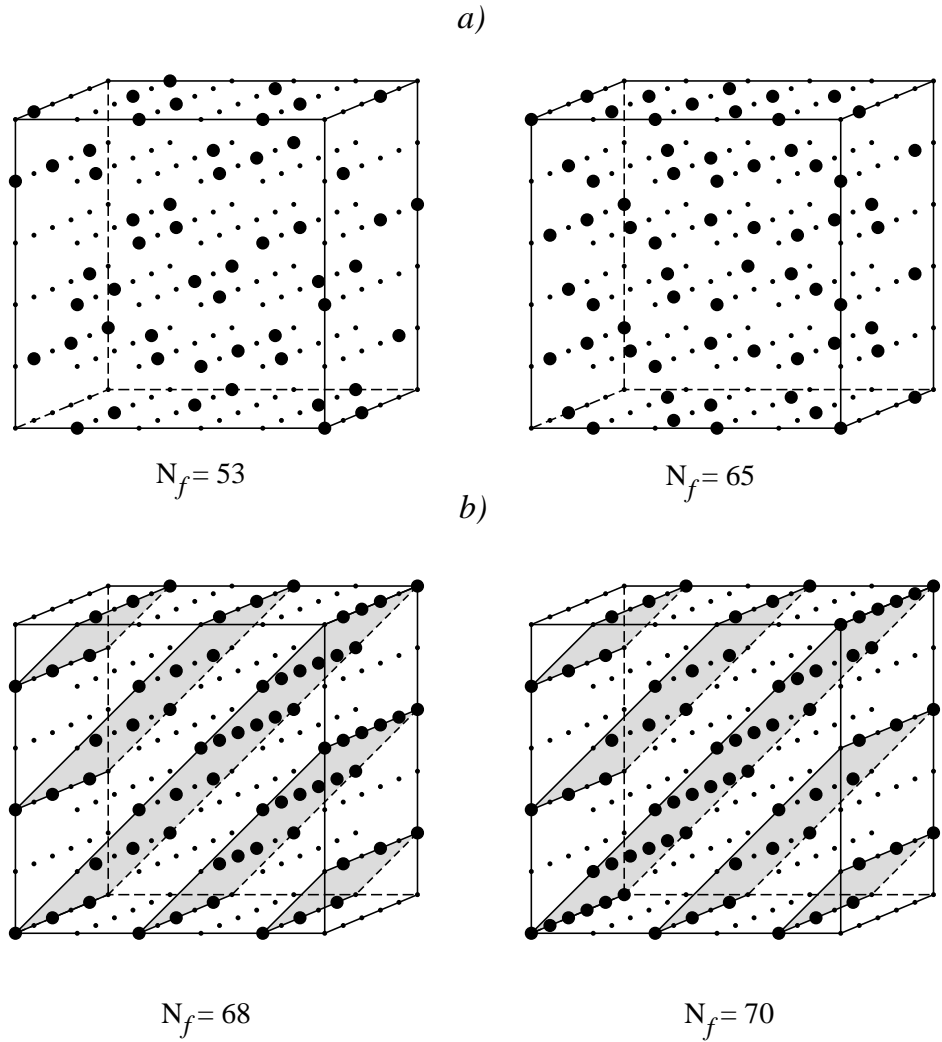}
\caption{ 
The ground state configurations for intermediate $f$-electron
concentartions. (a) The formation of the chessboard structure.  
(b) The examples of ground-state configurations that can be considered
as mixtures of configuration types with smaller $n_f$ 
($U=1$, $L=6\times 6\times 6$).
}
\label{fig3}
\end{figure}

\newpage
\begin{figure}[ht]
\hspace{1cm}
\includegraphics[angle=0,width=14cm,scale=1]{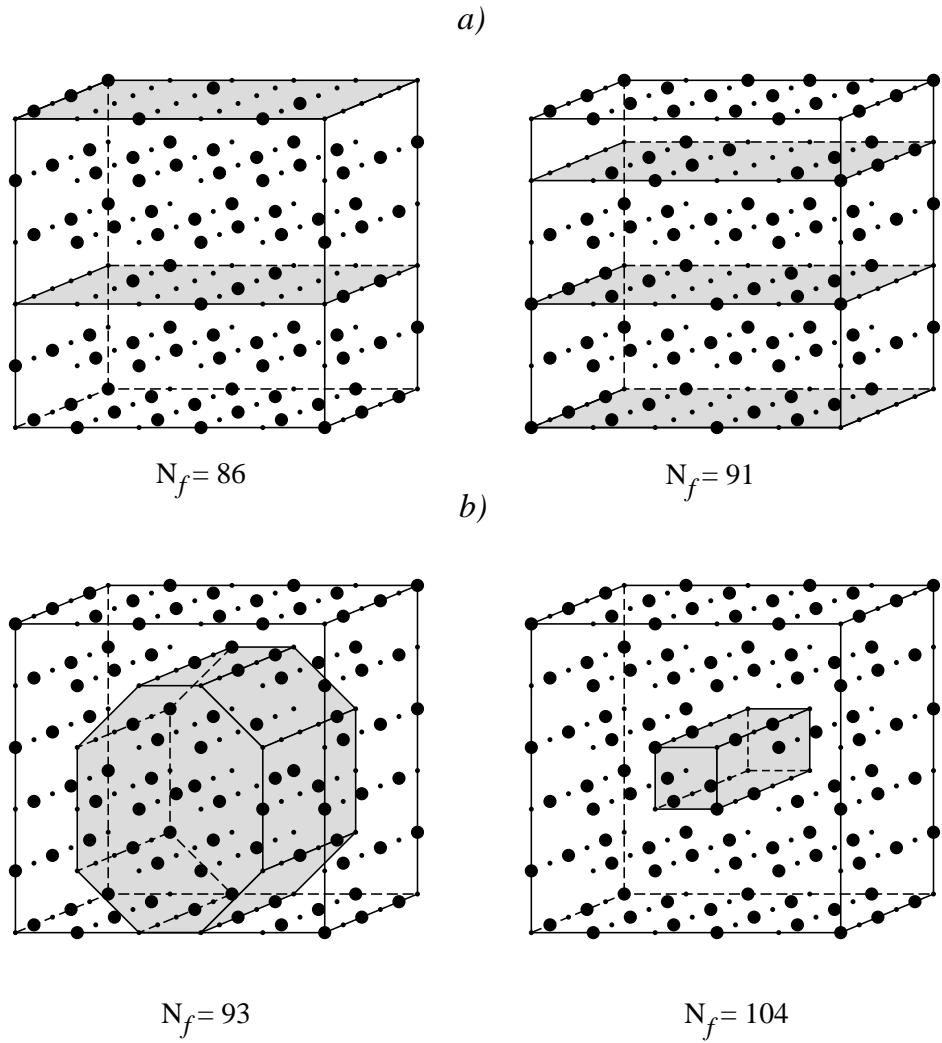}
\caption{ 
Examples of an incomplete chessboard structure obtained for $U=1$ 
and $L=6\times 6\times 6$. 
(a) The chessboard structure is fully developed in some regions (planes) 
that are separated by planes with incomplete developed 
chessboard structure. (b) Both regions with complete and incomplete 
chessboard structure have the three-dimensional character.
}
\label{fig4}
\end{figure}

\end{document}